# ВЛИЯНИЕ ПОГЛОЩЕНИЯ ИЗЛУЧЕНИЯ МИКРОЧАСТИЦАМИ НА СКОРОСТЬ ПЛАМЕНИ И РЕЖИМЫ ГОРЕНИЯ

*М.Ф. Иванов [a], А.Д. Киверин [a], М.А. Либерман [b]\**

[a] *Объединеннный Институт Высоких температур РАН,
125412 Москва, Россия*
[b] *Nordita, KTH Royal Institute of Technology and Stockholm University Roslagstullsbacken 23,
10691 Stockholm, Sweden*

Тепловое излучение горячих продуктов горения практически не влияет на распространение пламени в газовой среде. Мы рассматриваем иную ситуацию, когда даже небольшая концентрация взвешенных в газе микрочастиц поглощает тепловое излучение и, передавая его газу, нагревает газовую смесь перед фронтом волны горения. В зависимости от реактивности газовой смеси и величины нормальной скорости ламинарного пламени нагрев смеси перед фронтом пламени может приводить либо к умеренному в случае быстрого пламени, либо к значительному в случае медленного пламени увеличению скорости волны горения. Для достаточно медленного пламени перенос тепла излучением от продуктов горения может стать доминирующим механизмом по сравнению с обычной молекулярной теплопроводностью, определяющим структуру и скорость волны горения. Показано, что при неоднородном в пространстве перед фронтом пламени распределения частиц радиационный нагрев приводит к формированию градиента температуры, который в свою очередь может приводить к зажиганию различных режимов горения в зависимости от длины поглощения. В соответствии с градиентным механизмом Зельдовича при этом могут формироваться как режимы медленного горения, так и детонация. Зажигание различных режимов распространения волн горения в зависимости от величины длины поглощения излучения иллюстрируется на примере водород-кислородного пламени.

___________________________________________
\*E-mail: michael.liberman@nordita.org

# 1. ВВЕДЕНИЕ

Тепловое излучение продуктов горения высокой температуры, образующихся за фронтом пламени, практически не оказывает влияния на распространение пламени в газовой смеси. При нормальных условиях сечения рассеяния и поглощения излучения в газах крайне малы [1], а длина поглощения квантов в газовой смеси составляет десятки метров, так что величина поглощаемой энергии теплового излучения ничтожно мала. В случае волны горения, распространяющейся от закрытого конца трубы, тепловые потери от излучения продуктов горения незначительны по сравнению с потерями тепла на стенки трубы за счет теплопроводности. Ситуация существенно изменяется в присутствии даже относительно небольшой концентрации микрочастиц, которые поглощают излучение, нагреваются и нагревают окружающий газ. Например, для концентрации частиц микронного размера, $N_p \sim 10^7 \text{см}^{-3}$, длина поглощения излучения $L \approx 1/N_p \pi r_p^2$ порядка 1см, что определяет прогрев свежей смеси на масштабах порядка 1см перед фронтом пламени за счет поглощения теплового излучения от горячих продуктов горения микрочастицами. При этом, если массовая плотность частиц много меньше массовой плотности газа, $\varsigma = m_p N_p / \rho_g \ll 1$, то обратное влияние частиц на динамику формирующегося потока и динамику волны горения мало. Таким образом, именно радиационный прогрев определяет дальнейшую динамику фронта пламени, способствуя его ускорению. Следует также отметить, что уже при небольшой концентрации микрочастиц значительно возрастает светимость и оптическая толщина среды, так что излучение продуктов горения из-за фронта пламени можно считать равновесным излучением черного тела. Влияние излучения на режимы распространения волны горения представляет фундаментальный физический интерес для многих задач как классического, так и термоядерного горения. С практической точки зрения наибольший интерес представляют зажигание и динамика горения топливно-воздушных смесей с угольной пылью, горение в ракетных двигателях на твердом и жидком топливе, горение в дизельных двигателях и газовых турбинах. Горючие газовые смеси в запыленных помещениях типичны для угольных шахт и химического производства, поэтому исследование роли излучения при горении газовых смесей в присутствии микроскопических частиц представляет особый интерес в связи с проблемами безопасности. Учет излучения также может быть важен для понимания особенностей термоядерного горения в Белых карликах при образовании Сверхновых.

Динамика пламени в газовой смеси с частицами с учетом излучения исследовалась аналитически, используя асимптотические разложения для больших значений энергии активации в приближении одноступенчатой модели химической реакции в [2-5]. Распространение пламени в однородной газовой среде с однородным распределением инертных частиц с учетом и без учета переноса излучения исследовалось в связи с проблемами безопасности в угольных шахтах [6-11]. При этом основное внимание уделялось летучести угольной пыли в результате нагрева и практически важной с точки зрения безопасности возможности случайного возгорания.

В настоящей работе рассматривается эффект теплового радиационного нагрева перед фронтом пламени в двухфазной среде стехиометрической смеси водорода с кислородом или водорода с воздухом и взвешенных в ней химически инертных микрочастиц. Газовая смесь предполагается прозрачной для излучения, тогда как частицы поглощают излучение, нагреваются и посредством теплопроводности нагревают окружающую газовую смесь. Численное моделирование влияния излучения на динамику распространение пламени проводилось для полной системы уравнений Навье-Стокса для всех компонент участвующих в химических реакциях горения, с учетом вязкости, теплопроводности, диффузии и детальной кинетики химических реакций. Рассмотрены различные сценарии, связанные с поглощением излучения частицами перед фронтом пламени, в зависимости от степени однородности распределения частиц и, как следствие, от степени нагрева газовой смеси перед фронтом пламени и от нормальной ламинарной скорости пламени. При однородном пространственном распределении частиц, поглощение излучения и нагрев газовой смеси перед фронтом приводят к росту скорости распространения пламени. Эффект нагрева смеси перед фронтом и увеличения скорости волны горения тем больше, чем меньше нормальная скорость пламени и чем выше температура продуктов горения. Более того, в случае пламени с достаточно малой нормальной скоростью, эффективное время нагрева частиц и газовой смеси излучением может быть достаточным для нагрева газовой смеси до температуры зажигания (температура перехода индукционной эндотермической стадии реакции к быстрой экзотермической фазе). В этом случае перенос энергии тепловым излучением в результате поглощения излучения частицами и дальнейшей передача энергии от частиц газу, может стать доминирующим процессом по сравнению с классическим механизмом распространения ламинарного пламени за счет молекулярной теплопроводности. При этом структура фронта пламени меняется, заметно увеличивается ширина фронта и скорость распространения волны горения. При неоднородном распределении микрочастиц в объеме поглощение излучения локализовано в областях с более высокой концентрацией частиц, в результате чего смесь перед фронтом пламени нагревается неоднородно. Если нормальная скорость пламени мала так, что фронт набегающей волны горения не успевает поглотить область нагреваемой излучением смеси за время ее прогрева до величины зажигания реакции, то результатом неоднородного нагрева газовой смеси будет зажигание новой волны горения по градиентному механизму

Зельдовича [12, 13]. В зависимости от величины температурного градиента, определяемого в основном длиной поглощения излучения, может сформироваться либо волна медленного горения, либо детонация. В принципе, детонация может зажечься и в случае начального однородного распределении частиц для пламени с небольшой скоростью, в медленно реагирующей смеси. Однако, это было бы возможно только для довольно больших длин поглощения, поскольку минимальный масштаб температурного градиента, требуемый для зажигания детонации в более медленно реагирующей смеси по сравнению с водород кислородной, на порядки больше чем для водород кислородной смеси [14, 15]. В настоящей работе зажигание различных режимов горения перед фронтом пламени, связанное с поглощением излучения, продемонстрировано на примере быстрого пламени, распространяющегося по водород-кислородной смеси с неоднородным распределением микрочастиц, таким, что концентрация частиц мала вблизи начального положения фронта пламени и растет по мере удаления от него. Расстояние между областью существенного роста концентрации частиц и начального положения фронта пламени выбиралось таким образом, что температура смеси в области поглощения излучения успевает вырасти до температуры зажигания прежде, чем пламя достигнет этой области.

## 2. ПОСТАНОВКА ЗАДАЧИ

Мы рассматриваем одномерную задачу распространения плоского фронта пламени в двухфазной газовой стехиометрической смеси водорода с кислородом и взвешенных в ней нейтральных частиц микронного размера. Волна горения распространяется в канале от закрытого конца к открытому концу. Для газовой фазы решается система уравнений Навье-Стокса для сжимаемой среды с учётом вязкости, теплопроводности, многокомпонентной диффузии, энерговыделения в химических реакциях и обмена импульсом и энергией с частицами.

$$\frac{\partial \rho}{\partial t} + \frac{\partial (\rho u)}{\partial x} = 0, \tag{1}$$

$$\frac{\partial Y_i}{\partial t} + u\frac{\partial Y_i}{\partial x} = \frac{1}{\rho}\frac{\partial}{\partial x}\left(\rho D_i \frac{\partial Y_i}{\partial x}\right) + \left(\frac{\partial Y_i}{\partial t}\right)_{ch}, \tag{2}$$

$$\rho\left(\frac{\partial u}{\partial t} + u\frac{\partial u}{\partial x}\right) = -\frac{\partial P}{\partial x} + \frac{\partial \sigma_{xx}}{\partial x} - \rho_p \frac{(u - u_p)}{\tau_{St}}, \tag{3}$$

$$\rho\left(\frac{\partial E}{\partial t} + u\frac{\partial E}{\partial x}\right) = -\frac{\partial (Pu)}{\partial x} + \frac{\partial}{\partial x}(\sigma_{xx} u) + \frac{\partial}{\partial x}\left(\kappa(T)\frac{\partial T}{\partial x}\right) +$$

$$+ \sum_k h_k \left(\frac{\partial}{\partial x}\left(\rho D_k(T)\frac{\partial Y_k}{\partial x}\right)\right) + \rho \sum_k h_k \left(\frac{\partial Y_k}{\partial t}\right)_{ch} - \rho_p u_p \frac{(u - u_p)}{\tau_{St}} - \rho_p c_{P,p} Q, \tag{4}$$

$$P = R_B T n = \left(\sum_i \frac{R_B}{m_i} Y_i\right) \rho T = \rho T \sum_i R_i Y_i, \tag{5}$$

$$\varepsilon = c_v T + \sum_k \frac{h_k \rho_k}{\rho} = c_v T + \sum_k h_k Y_k, \tag{6}$$

$$\sigma_{xx} = \frac{4}{3}\mu\left(\frac{\partial u}{\partial x}\right) \tag{7}$$

В уравнениях (1-7) приняты стандартные обозначения: $P$, $\rho$, $u$ давление, массовая плотность и скорость газовой фазы, $Y_i = \rho_i / \rho$ - массовая концентрация компонент смеси, $E = \varepsilon + u^2/2$ - полная плотность энергии, $\varepsilon$ -плотность внутренней энергии, $R_B$ - универсальная газовая постоянная, $m_i$ - молярная масса i-компоненты смеси, $R_i = R_B / m_i$, $n$ - молярная плотность газовой смеси, $\sigma_{ij}$ - тензор вязких напряжений, $c_v = \sum_i c_{vi} Y_i$ - теплоемкость при постоянном объеме, $c_{vi}$ - теплоемкость i-компоненты газовой фазы, $h_i$ - энтальпия образования i-компоненты, κ(T) и μ(T) коэффициенты теплопроводности и вязкости, $D_i(T)$ - коэффициент диффузии i-компоненты, $(\partial Y_i / \partial t)_{ch}$ - изменение концентрации i-компоненты в химической реакции, $\rho_p = m_p N_p$ – массовая плотность частиц, $N_p$ – концентрация частиц, $u_p$ – скорость частиц, $\tau_{St} = m_p / 6\pi\mu r_p$ - характерное Стоксовское время для сферических частиц радиуса $r_p$ и массы $m_p$, $\rho_p C_{P,p} Q$ – изменение тепловой энергии в результате межфазного теплообмена, $c_{P,p}$ – удельная теплоемкость частиц. Уравнение состояния смеси определялось по зависимости теплоемкостей и удельных энтальпий отдельных компонент смеси от температуры, которые определялись из теплофизических таблиц JANAF путем интерполяции полиномами пятого порядка [16], коэффициенты переноса рассчитывались на основании кинетической теории газов [17].

Значения концентраций компонент $Y_i$ реагирующей газовой смеси определялись из решения системы уравнений химической кинетики

$$\frac{dY_i}{dt} = F_i(Y_1, Y_2, ... Y_N, T), \quad i = 1, 2, ... N, \tag{8}$$

где правая часть уравнений (8) содержит скорости химических реакций для компонент $H_2$, $O_2$, $H$, $O$, $OH$, $H_2O$, $H_2O_2$, $HO_2$, зависящие от температуры по закону Аррениуса в стандартном виде. Для моделирования химической кинетики реакций использовались стандартные редуцированные модели кинетики водород-кислородной и водород-воздушной смеси с энергиями активации для элементарных реакций, приведенными в [18].

Фаза взвешенных в газе микрочастиц рассматривается в континуальном приближении, в котором динамика частиц определяется уравнениями переноса, подобным уравнениям газодинамики. Континуальное описание для ансамбля частиц возможно, если соответствующий масштаб, характеризующий частицы пренебрежимо мал по сравнению с характерным масштабом изменения параметров течения, но достаточно большой, чтобы количество частиц на таком масштабе обеспечивало корректное описание их статистически осредненных параметров. Для частиц размерами порядка 1мкм и достаточно малой объёмной концентрации, континуальное описание динамики частиц является оправданным. Взаимодействием частиц друг с другом при рассматриваемых здесь концентрациях и размерах частиц можно пренебречь, ограничившись только их взаимодействием с газовой фазой, описываемым стоксовской силой трения. Тогда система уравнений для частиц:

$$\frac{\partial N_p}{\partial t} + \frac{\partial (N_p u_p)}{\partial x} = 0 \tag{9}$$

$$\left(\frac{\partial u_p}{\partial t} + u_p \frac{\partial u_p}{\partial x}\right) = \frac{(u - u_p)}{\tau_{St}}, \tag{10}$$

$$\left(\frac{\partial T_p}{\partial t} + u_p \frac{\partial T_p}{\partial x}\right) = Q - \frac{2\pi r_p^2 N_p}{c_{P,p} \rho_{p0}} \left(4\sigma T_p^4 - q_{rad}\right). \tag{11}$$

Где $T_p$ – температура частиц, $2\pi r_p^2 N_p \left(4\sigma T_p^4 - q_{rad}\right)$ поток теплового излучения, поглощаемый и переизлучаемый частицами.

$$Q = (T_p - T)/\tau_{pg}, \tag{12}$$

где $\tau_{pg} = 2r_p^2 c_{P,p} \rho_{p0} / 3\kappa Nu$ - характерное время передачи тепла от частиц газовой фазе, $\rho_{p0}$ - массовая плотность частицы, $Nu \approx 2$ - число Нуссельта [19].

Для одномерной плоской задачи, уравнение переноса теплового излучения в диффузионном приближении имеет вид [1, 20]:

$$\frac{d}{dx}\left(\frac{1}{\chi}\frac{dq_{rad}}{dx}\right) = -3\chi\left(4\sigma T_p^4 - q_{rad}\right), \tag{13}$$

где $\chi = 1/L = \pi r_p^2 N_p$ - коэффициент поглощения теплового излучения, где $L = 1/\pi r_p^2 N_p$ - средняя длина поглощения излучения. В такой постановке задачи тепловое излучение приходит от частиц нагретых горячими продуктами горения за фронтом пламени. При этом оптическая толщина продуктов горения за фронтом пламени достаточно велика, так что поток излучения, исходящий с фронта пламени, равен равновесному излучению с черного тела, $q_{rad}(x = X_f) = \sigma T_b^4$, где $\sigma = 5.6703 \cdot 10^{-5}$ эрг/см$^2$ K$^4$ – постоянная Стефана-Больцмана, $T_b \approx 3000 K$ температура продуктов горения водород-кислородной смеси.

Для численного решения системы уравнений (1)-(7), (9)-(11) использовался эйлерово-лагранжев метод, известный как метод «крупных частиц» (МКЧ) [21]. В настоящей работе использовался модифицированный метод, имеющий точность второго порядка по пространству, который показал хорошее согласие с экспериментами при исследованиях динамики пламени в широких трубах, для численного моделировании стука в двигателях внутреннего сгорания [22, 23], и исследовании ускорения водород-кислородного пламени в трубах и перехода горения в детонацию [24-28]. Система уравнений химической кинетики решалась методом Гира [29]. Детали используемого в работе численного метода и исследование сходимости решений приведены в Приложении.

## 3. ПЕРЕНОС ТЕПЛОВОГО ИЗЛУЧЕНИЯ И УСКОРЕНИЕ ПЛАМЕНИ ПРИ ОДНОРОДНОМ НАЧАЛЬНОМ РАСПРЕДЕЛЕНИИ ЧАСТИЦ

Рассмотрим распространение пламени в стехиометрической водород-кислородной смеси с начальным пространственно однородным распределением микрочастиц. Частицы в газе перед фронтом пламени поглощают тепловое излучение, выходящее из фронта, нагреваются и в свою очередь передают тепло окружающему газу. При распространении от закрытого конца трубы, перед фронтом пламени формируется поток газа, вызванный расширением продуктов горения. Скорость такого потока $(\theta-1)U_f$, при том, что скорость пламени в лабораторной системе равна $\theta U_f$, где $\theta$ отношение плотностей несгоревшей и сгоревшей смесей. Фронт пламени движется со скоростью $U_f$ относительно несгоревшей смеси, т.е. с той же скоростью нагретый излучением газ поступает во фронт пламени. Стационарное решение достигается, когда во фронт поступает максимально прогретый газ, т.е. когда фронт пламени поглощает смесь на расстоянии порядка длины поглощения излучения. Таким образом, стационарное решение устанавливается за время, когда фронт пламени проходит расстояния порядка $L/U_f$. Полученные в расчетах изменения во времени температуры перед фронтом пламени за счет поглощения излучения для разных длин поглощения показаны на Рис.1 для частиц радиуса $r_p = 0.75$ мкм и концентраций частиц $N_p = 5.7\cdot10^7; \; 2.85\cdot10^7; 1.4\cdot10^7$ см$^{-3}$, соответствующих длинам поглощения излучения L=1, 2, 4см. Динамические параметры в расчетах для массовой плотности частиц $\rho_{p,0} = 1$ г/см$^3$: $\varsigma = \rho_p/\rho = 0.2; \; 0.1$ и $0.05$, $\tau_{St} = 2r_p^2\rho_{p0}/9\nu_g\rho = 3$мкс, $\tau_{pg} \approx 1$ мкс, и $\tau_{gp} = \tau_{pg}\dfrac{\rho c_{V,g}}{\rho_p c_{P,p}} = (c_{V,g}/c_{P,p})/\varsigma$ порядка нескольких мкс. Поскольку $\varsigma = \rho_p/\rho \ll 1$, то передача импульса от частиц газу и обратное влияние частиц на динамику волны горения мало. В виду этого в отсутствии радиационного переноса тепла скорость пламени была бы практически одинакова для всех рассмотренных вариантов, и изменения в динамике пламени определяется в основном радиационным нагревом частиц и газовой смеси перед фронтом пламени. Для малых значений параметра $\varsigma \ll 1$, при $\varsigma(c_{p,p}/c_{V,g}) \ll 1$, уменьшение адиабатической температуры продуктов горения, в смеси разбавленной частицами, также относительно мало. Неравенство $\varsigma \ll 1$ полезно также записать в виде $L \gg r_p(\rho_{p0}/\rho_g)$, из которого следует естественное ограничение применимости рассматриваемой модели для длины поглощения излучения при данных размерах и плотности частиц.

Максимальную температуру в смеси перед фронтом пламени, достигаемую в процессе установления стационарного режима можно оценить из уравнения

$$\rho_p c_{p,p} \frac{dT_p}{dt} = \sigma T_b^4 \exp\left(-\frac{x - U_f t}{L}\right) \pi r_p^2 N_p - \frac{\rho_p c_{p,p}}{\tau_{pg}}(T_p - T). \qquad (14)$$

Учитывая, что временные масштабы обмена импульсом и энергией между газовой фазой и взвешенными микрочастицами малы по сравнению с характерным временем распространения волны горения, $t_f = L_f / U_f$ где $L_f$ – ширина фронта пламени, можно пренебречь различием в скорости и температуре газовой и дисперсной фаз, и обозначая $\xi = \tau_{gp}/\tau_{pg} = c_{V,g}/\varsigma c_{P,p}$, переписать (14) в виде

$$\rho_p c_{p,p}(1+\xi)\frac{dT}{dt} = \sigma T_b^4 \frac{1}{L}\exp\left(-\frac{x - U_f t}{L}\right) \qquad (15)$$

Уравнение (15) дает оценку максимального нагрева среды непосредственно перед фронтом пламени при установлении стационарного течения, т.е. температуру, которая устанавливается за время $L/U_f$.

$$\Delta T = \sigma T_b^4 \frac{1}{U_f} \frac{(1-e^{-1})}{\rho_p c_{P,p}(1+\xi)} \approx 0.63 \frac{\sigma T_b^4}{(\rho_p c_{P,p} + \rho c_{V,g})U_f}, \qquad (16)$$

При оценке максимальной температуры нагрева, даваемой формулой (16), нужно учитывать, что в смеси разбавленной нейтральными частицами температура за фронтом пламени несколько ниже адиабатической, порядка $T_b \approx (2900 \div 3000)K$. Поскольку фронт пламени имеет конечную толщину, а не является поверхностью разрыва, а кванты излучения, выходящие с поверхности, рождаются в слое толщиной порядка длины поглощения, то в (16) следует брать более низкую эффективную температуру излучения. Кроме того, формула (16) не учитывает, что время прогрева уменьшается с увеличением скорости пламени примерно на 20%. С учетом сделанных замечаний, оценка для повышения температуры за счет радиационного нагрева хорошо согласуется с результатами численных расчетов. Полученные в численных расчетах температурные профили, установившиеся перед фронтом пламени в стационарном режиме для тех же значений длин поглощения и параметров, что и на Рис.1, показаны на Рис.2 и хорошо согласуются с оценкой даваемой уравнением (16).

Следует отметить, что в соответствии с формулой (16) и проведенными расчетами максимальная температура радиационного нагрева почти не зависит от длины поглощения излучения. Это связано с тем, что хотя локальный нагрев частицы, движущейся в потоке перед фронтом пламени, больше при меньших длинах поглощения, однако при больших длинах поглощения время нагрева частицы излучением до установления стационарного режима больше, что компенсирует меньший локальный нагрев. Более высокая температура вблизи фронта пламени в стационарном режиме, устанавливающаяся в случае большей длины поглощения (Рис.1), обусловлена также тем, что в вариантах с меньшей длиной поглощения больше концентрация нейтральных частиц, и, следовательно, больше разбавление смеси и несколько ниже температура излучающей смеси за фронтом пламени. Согласно (16) время радиационного нагрева тем больше, чем меньше нормальная скорость ламинарного пламени в газовой смеси, и чем выше максимальная температура достигаемая в процессе радиационного нагрева. Увеличение температуры газовой смеси перед фронтом пламени на масштабе длины поглощения излучения L приводит к росту скорости распространения пламени. Скорость горения при повышенной температуре перед пламенем возрастает по сравнению со скоростью горения в нормальных условиях, и пламя ускоряется за счет переноса теплового излучения в дисперсной среде. Рост скорости горения для различных длин поглощения, вычисленный для условий Рис. 2, показан на Рис.3.

Интересно отметить, что в классической теории горения ширина фронта ламинарного пламени тем меньше, чем больше скорость пламени [30, 31]. По порядку величины скорость $U_f$ и ширина фронта $L_f$ ламинарного пламени связаны соотношением $L_f U_f \approx \chi$, где $\chi_g = \kappa_g / \rho c_{p,g}$, коэффициент температуропроводности. На рис. 4 показано изменение ширины фронта пламени, определяемой как $L_f = (T_b - T_f)/\max(\nabla T)$, где $T_b$ – температура продуктов горения, $T_f$ – температура перед фронтом пламени. Видно, что ширина фронта пламени увеличивается примерно на 15-20%, что сравнимо с увеличением скорости волны горения в результате радиационного нагрева и обусловлено радиационным переносом тепла по рассматриваемому механизму.

Для более медленно реагирующих смесей, в которых нормальная скорость ламинарного пламени значительно меньше, температура смеси перед фронтом пламени в результате радиационного нагрева может возрасти до температуры зажигания. Например, для пламени в смеси метан-воздух, $T_b \approx 2200K$, $U_f \approx 30$ см/сек, согласно (16) температура в результате радиационного нагрева должна была бы вырасти на $\Delta T \approx 2500K$. Еще больше радиационный нагрев перед фронтом пламени в смеси пропана с воздухом и при горении закиси углерода с кислородом $(CO + 0.5O_2)$. Очевидно, что столь большие значения температуры в смеси перед фронтом не имеют отношения к действительности. Задолго до того повышение температуры в несгоревшей смеси приведет к зажиганию реакции в газе перед фронтом пламени. При этом механизмом распространения волны горения становится уже не перенос тепла молекулярной теплопроводностью, а перенос тепла за счет поглощения теплового излучения частицами. Необходимым для этого условием является нагрев излучением частиц и газовой фазы до температуры зажигания горючей смеси. Можно оценить ширину фронта и скорость волны горения в случае, когда радиационный перенос тепла за счет поглощения излучения частицами становится доминирующим механизмом. Для ламинарного пламени в газовой смеси распространяющегося благодаря молекулярной диффузии тепла (или радикалов) оценка для ширины фронта и скорости пламени может быть получена, учитывая, что характерное диффузионное время должно быть больше характерного времени выделения тепла в химической реакции $\tau_R$, чтобы диффузия малых возмущений исходящих от пламени не гасила горение. Отсюда, из соображений размерности [32] следует

$$L_f \propto \sqrt{\chi_g \tau_R}, \; U_f \propto \sqrt{\chi_g / \tau_R}, \qquad (17)$$

где $\tau_R$ характерное время реакции.

В случае, когда радиационный перенос тепла за счет поглощения излучения частицами становится доминирующим механизмом, ширина фронта образующейся волны горения по порядку величины равна длине поглощения излучения. Можно сказать, что при доминирующем радиационном переносе тепла реакция инициируется на масштабе порядка длины поглощения излучения, что определяет распространение реакции в пространстве. Напомним, что при молекулярном переносе тепла ширина зоны реакции, много меньше ширины фронта пламени. При доминирующем радиационном переносе тепла скорость распространения волны горения можно по порядку величины оценить как

$$U_{f,rad} \propto (L/\tau_R) = (L/L_f)U_{f0}. \tag{18}$$

Для $L \gg L_f$, скорость такой волны горения, $U_{f_{rad}} \propto (L/L_f)U_{f0} \gg U_{f0}$. Такая волна горения будет выглядеть как последовательность тепловых взрывов, сопровождающихся сильным ростом давления. Можно предположить, что катастрофы в угольных шахтах, происходящие при взрывах шахтного газа (метана) при его случайном воспламенении не обязательно обусловлены возникновением детонации, а связаны с переходом обычного режима горения к описанному выше горению в газовой смеси с угольной пылью, когда механизмом распространения волны горения становится перенос тепла за счет поглощения теплового излучения мелкими частицами угольной пыли.

Следует отметить, что возможен другой сценарий "радиационного теплового взрыва". В условиях, когда перенос тепла излучением становится доминирующим процессом, в принципе, возможно зажигание дефлаграции, или детонации по механизму Зельдовича [12] в зависимости от величины градиента температуры образовавшегося в области поглощения излучения перед фронтом пламени. При этом необходимо, чтобы время индукции было много меньше характерного гидродинамического временем ($L/U_f$), для того чтобы ускоряющееся первичное пламя не успело поглотить нагретую впереди область, где происходит зажигание реакции. Например, для пламени в водород-кислородной смеси при начальном давлении (и плотности) в несколько раз меньшем атмосферного давления, возможен нагрев тепловым излучением, поглощаемым частицами, при котором температура в области поглощения становится больше температуры зажигания, и перенос тепла излучением станет доминирующим процессом. В этом случае, как и для достаточно медленного пламени при нормальных условиях и при достаточно больших длинах поглощения излучения перед фронтом мог бы образоваться достаточно пологий температурный градиент для зажигания детонации. Однако при давлениях меньше атмосферного и для относительно медленно реагирующих смесей масштаб температурного градиента, при котором зажигается детонация значительно возрастает. Например минимальный масштаб линейного температурного градиента, зажигающего детонацию в водород воздухе или в водород кислородной смеси при начальном давлении 0.1атм, около одного метра по сравнению с несколькими сантиметрами для водород кислородной смеси при начальном давлении 1атм. Численное моделирование даже одномерной задачи в этом случае требует больших вычислительных ресурсов. Поэтому, в следующем разделе мы продемонстрируем аналогичный сценарий, реализуемый при неоднородном начальном распределении частиц.

Следует также отметить, что в рассматриваемой ситуации отсутствует локальное термодинамическое равновесие с тепловым излучением, и, следовательно, неприменимо приближение лучистой теплопроводности. Излучение может быть в термодинамическом равновесии с веществом, например в условиях термоядерного горения в звездах, где среда оптически толстая и температура мало меняется на длине поглощения. В этом случае конкуренция переноса тепла излучением с теплопроводностью вещества можно оценить, сравнивая коэффициенты лучистой теплопроводности $\kappa_{rad} = 16\sigma L T^3 / 3$ и теплопроводности среды. Например, для термоядерного горения Белых карликов при взрыве Сверхновых коэффициент электронной теплопроводности вырожденного электронного газа [33], $\kappa_e \propto 10^{10} T/K$ сравним по величине с коэффициентом лучистой теплопроводности [34]. При этом интересно отметить, что в случае невырожденного электронного газа, электронная теплопроводность ($\kappa_e = (1.3/\Lambda Z) \cdot 10^{11} T_e^{5/2}$) будет всегда больше лучистой теплопроводности даже при очень высоких температурах, соответствующих температурам при термоядерном горении.

## 4. ПЛАМЯ В СМЕСИ С НЕОДНОРОДНЫМ РАСПРЕДЕЛЕНИЕМ ЧАСТИЦ: ЗАЖИГАНИЕ ГОРЕНИЯ И ДЕТОНАЦИИ

Мы проиллюстрируем зажигание различных режимов горения на примере распространения пламени в водород кислородной газовой смеси с неоднородным пространственным распределением микрочастиц. Легко видеть, что для того чтобы температура частиц и газа перед фронтом пламени выросла за счет поглощения излучения до температуры зажигания, $T \approx 1050K$ при $P_0 = 1.0$ атм, время нагрева излучением должно быть примерно $t_{rad} \approx 0.001$ с. За это время волна горения в водород-кислородной смеси ($U_f \approx 1200$ см/с) проходит расстояние порядка 1см. Если при этом концентрация микронных частиц непосредственно перед фронтом пламени $N_p \sim 10^6$ см$^{-3}$, а длина поглощения излучения L>>10см, а дальше от фронта концентрация частиц возрастает, образуя более плотное облако, то излучение от фронта пламени доходит почти без потерь до области с более высокой концентрацией частиц, где оно поглощается и нагревает смесь прежде чем волна горения достигнет границы плотного облака частиц. Соответствующее распределение концентрации частиц и положение фронта пламени показаны схематически на Рис. 5.

Градиент температуры в газе, образующийся в результате поглощения излучения частицами и нагрева газа определяется в основном длиной поглощения излучения и расширением газа в процессе нагрева. Если при этом поглощение излучения происходит достаточно далеко от фронта пламени, так что максимальная температура в вершине температурного градиента вырастает до температуры зажигания, то в области градиента температуры зажигается новая волна горения по механизму Зельдовича, которая в зависимости от величины температурного градиента [14, 15] может сформироваться либо в режиме медленного горения, либо детонации.

Нагрев среды излучением ведет к локальному расширению газа в области поглощения излучения и, как следствие, к перераспределению плотности газа и частиц вблизи границы облака. Поскольку характерное время радиационного нагрева смеси до температуры воспламенения (~0.001с), много больших характерного акустического времени, давление в области радиационного нагрева выравнивается, и градиент температуры образуется при постоянном давлении, близком к начальному. На Рис. 6 показаны результаты численного моделирования эволюции во времени профилей температуры в газовой смеси и плотности частиц в ходе радиационного нагрева для начального распределения плотности частиц радиуса $r_p = 1$ мкм в виде ступеньки с нарастанием концентрации частиц до значения $N_p = 2.5 \cdot 10^7$ см$^{-3}$. Плотность частиц между фронтом пламени и левой, обращённой в сторону фронта пламени, границей облака частиц должна быть, по крайней мере, на порядок меньше максимальной плотности, так что длина поглощения в этой области много больше расстояния от фронта пламени до левой границы облака частиц. Вместе с тем следует отметить, что даже такая, относительно небольшая плотность частиц в области пламени значительно увеличивает светимость продуктов горения и их оптическую толщину. Поэтому в численной модели излучение, выходящее из фронта пламени, принимается равным равновесному излучению черного тела.

На Рис.7 показана эволюция во времени профилей плотности, температуры и давления, начиная с момента $t_0 = 900$ мкс, когда максимальная температура газа выросла до значения температуры зажигания (температуры, при которой воспламенение происходит на меньших временных масштабах по сравнению с дальнейшим разогревом среды излучением). Характерный пространственный масштаб температурного градиента, приведенного на Рис. 7, $(T^* - T_0)/|dT/dx| \approx 1$ см близок к длине поглощения излучения частицами $L = 1/\pi r_p^2 N_p \approx 1.2$ см. Согласно классификации режимов горения инициируемых градиентным механизмом Зельдовича, полученной с учетом детальной химической кинетики [14, 15], при начальном давлении 1 атм такой температурный градиент в водород-кислородной смеси инициирует волну дефлаграции. Эволюция во времени профилей температуры на Рис. 7 показывает развитие спонтанной волны горения на сформированном градиенте температуры, которая затем переходит в нормальную волну дефлаграции. Пунктирная линия на верхней части рисунка показывает профиль начальной плотности частиц, и сплошная линия показывает распределение концентрации частиц в момент времени $t_0 = 900$ мкм, когда сформировался соответствующий температурный градиент в газовой смеси с температурой $T^* = 1050$ К в вершине градиента. Профили давления на Рис.7 показывают относительно небольшое повышение давления при воспламенении, развитии спонтанной волны горения, и формировании дефлаграции.

Для случая более низкой плотности частиц, и соответственно, большей длины поглощения излучения, начальный градиент температуры будет более пологим, и результатом будет образование "быстрой" волны дефлаграции, распространяющейся позади уходящей слабой ударной волны [14, 15]. Такая ситуация возникает в случае диффузной границы, когда концентрация частиц линейно спадает на левой границе облака частиц, обращенной в сторону пламени (Рис.8).

На Рис. 8 представлены результаты расчетов для случая, когда концентрация частиц линейно спадает на масштабе 1см при максимальной плотности частиц такой же, как на Рис. 7. Начальная концентрация частиц линейно спадает на левой границе облака, и показана штриховой линией. Концентрации частиц в момент $t_0 = 1650$ мкм перед зажиганием показана сплошной линией. Временная эволюция профилей температуры на Рис. 8 показывает развитие спонтанной волны горения и формирование быстрой волны дефлаграции за отходящей ударной волной.

Для образования детонации начальный температурный градиент должен быть еще более пологим, так чтобы скорость спонтанной волны горения, бегущей вдоль градиента температуры, в точке минимума была больше локальной скорости звука, что обеспечивает положительную обратную связь спонтанной волны реакции с образующимся импульсом давления, их взаимное усиление, и в конечном счете образование сильной ударной волны и детонации. Согласно [14, 15] для образования детонации на линейном температурном градиенте при температуре $T^* = 1050$ К в верхней точке градиента и при нормальных условиях, $P_0 = 1$ атм, $T_0 = 300$ К, минимальный масштаб градиента составляет около 20см. Такой температурный градиент образуется в случае диффузной границы, когда концентрация частиц на границе возрастает линейно на масштабе 10см до $N_p = 2.5 \cdot 10^7$ см$^{-3}$ как показано на верхней части Рис. 9 пунктирной линией. Сплошная линия здесь показывает распределение плотности частиц, получающееся в результате нагрева и расширения газа на момент $t_0 = 4980$ мкс, когда происходит зажигание реакции в верхней точке градиента. Временная эволюция профилей температуры и давления на Рис. 9 показывает развитие спонтанной волны горения, распространяющейся вдоль градиента температуры, образование ударной волны, и переходный процесс взаимодействия спонтанной волны горения с ударной волной, когда в результате положительной обратной связи происходит ускорение реакции, усиление ударной волны и образование волны детонации.

## 5. ЗАКЛЮЧЕНИЕ

В настоящей работе исследовано влияние теплового излучения продуктов горения на режимы распространения пламени в двухфазной среде, состоящей из газовой горючей смеси и взвешенных в ней нейтральных микрочастиц, поглощающих излучение. На примере волны горения в водород-кислородной смеси и более медленного пламени в водородно-воздушной смеси показано, что нагрев газовой смеси перед фронтом пламени за счет поглощения излучения микрочастицами относительно небольшой массовой концентрации приводит к росту скорости распространения волны горения. Вообще говоря, нагрев и увеличение скорости при прочих одинаковых условиях тем сильнее, чем меньше нормальная скорость ламинарного пламени в той же самой газовой смеси без частиц. В случае достаточно небольшой нормальной скорости ламинарного пламени перенос тепла за счет поглощения излучения и нагрева газовой фазы частицами может стать доминирующим процессом по сравнению с обычной молекулярной теплопроводностью. В этом случае возможно не только значительное увеличение скорости распространения волны горения, сопровождающееся изменением структуры фронта волны, но также зажигание перед фронтом пламени новых волн горения, которые в зависимости от длины поглощения излучения могут развиваться либо в режиме медленного горения, либо в детонационном режиме. В настоящей работе такой сценарий зажигания дефлаграции и детонации продемонстрирован для случая пламени, распространяющегося в двухфазной смеси с неоднородным распределением поглощающих излучение частиц, когда нагрев смеси до температуры зажигания происходит до того, как исходное пламя дойдет до области, где концентрация частиц достаточно велика для поглощения излучения.

Подытоживая изложенное можно сказать, что излучение горячих продуктов горения и поглощение излучения перед фронтом пламени в присутствии даже небольшой концентрации микрочастиц является существенным фактором, влияющим на режим и интенсивность горения. Нужно отметить, что изложенные здесь результаты получены для одномерной модели. В зависимости от внешних факторов, геометрии сосуда, шероховатости стенок, гидродинамики течения, и т.п. вся картина явления существенно усложняется.

Рассмотренный в работе сценарий образования волны детонации в результате нагрева вещества излучением с формированием подходящего для инициирования детонации температурного градиента, может представлять интерес для теории термоядерного горения Сверхновых типа Ia. Известно, что регулярность и корреляция яркости кривых светимости Сверхновых типа Ia делают их уникальными объектами для измерений постоянной Хаббла и скорости расширения вселенной [35-37]. Однако, погрешность измерений постоянной Хаббла на сегодняшний день составляет 10%, и для улучшение точности измерений требуется лучшее теоретическое описание кривых светимости, для чего необходимы улучшенные теоретические модели горения Сверхновой типа Ia. Сверхновая типа Ia является результатом термоядерного горения углерода и/или кислорода Белого карлика. Кроме основных продуктов горения, $^{56}$Ni, $^{56}$Co, $^{56}$Fe, которые в результате бета-распада служат мощным источником гамма-излучения, который и обуславливает яркую вспышку Сверхновой, образуются также "промежуточные" элементы, такие как Mg, Si, S, Ca, и др. Другой особенностью является наблюдаемый выброс вещества с поверхности на поздней стадии горения Сверхновой. Таким образом, модель горения Сверхновой должна объяснять: неполное выгорание и спектральное распределение промежуточных элементов и выброс вещества с поверхности. Численные модели, которые наилучшим образом воспроизводят спектр элементов в продуктах горения, известны как модели с "запаздывающей" детонацией, предложенные авторами работ [38, 39]. В этих моделях горения сверхновой происходит в режиме дефлаграции, которая вблизи последней стадии горения переходит в детонацию. Одной из трудностей моделирования физики горения сверхновой является огромное различие масштабов задачи. Начальный радиус звезды $10^8$ см, тогда как ширина фронта пламени меняется от $10^{-3}$ см в центре звезды, где плотность порядка $3 \cdot 10^9$ г/см$^3$ до примерно $10^{-2}$ см вдали от центра. Очевидно, что самосогласованное моделирование горения при таком различии в масштабах невозможно даже при использовании самых мощных компьютеров. Поэтому, несмотря на многочисленные попытки, как используя численное моделирование [38-44], так и аналитические модели [46, 47] до сих пор не удается убедительно объяснить механизм перехода медленного горения в детонацию в сверхновых, и переключение с одного режима на другой делается в численных моделях искусственно [44]. Сценарий зажигания детонации благодаря радиационному нагреву перед фронтом волны дефлаграции, аналогичный рассмотренному в настоящей работе, может быть полезен для понимания механизма перехода в детонацию при горении Сверхновой. В этом случае излучение поглощается самим веществом звезды. На стадии горения, когда волна горения отошла на достаточное расстояние от центра звезды, интенсивность излучения от переходов $Ni^{56} \rightarrow Co^{56} \rightarrow Fe^{56}$, произведенных уже в достаточно большом количестве, может, при поглощении внешних слоях звезды, создать температурный градиент, подходящий для инициирования детонации. Хотя, как было показано в [47], в вырожденном ферми газе детонация неустойчива относительно одномерных пульсаций во внутренних областях при плотностях больше ~ $2 \cdot 10^7$ г/см$^3$, при наличии температурного градиента детонация может окончательно сформироваться вблизи поверхности звезды.

**ПРИЛОЖЕНИЕ**

При проведении численного эксперимента сходимость решения играет такую же важную роль, как "воспроизводимость" опыта в физическом эксперименте. Поэтому была проведена большая серия тестов на сходимость решения для принципиально важных стадий развития процесса горения с переносом теплового излучения и поглощения его микрочастицами. При этом учитывались особенности развития процессов воспламенения и горения в двух постановках задачи: горение в двухфазной среде (химически активная газовая компонента, содержащая микрочастицы) и горение газовой смеси без частиц.

В случае распространения волны горения по двухфазной среде с равномерным распределением микрочастиц в объеме основным механизмом интенсификации горения является предварительный нагрев газовой компоненты по механизму, описанному в настоящей работе. Поэтому необходима проверка сходимости численных решений, как для нормальных начальных условий, так и для условий, с лучистым переносом тепла от горячих продуктов горения к свежей смеси перед распространяющимся фронтом пламени. Сходимость решений проверялась по сходимости для величин скорости горения, ширины фронта волны горения и характеристикам продуктов горения. На Рис. П1 показаны результаты тестов на сходимость решения для скорости ламинарного пламени, распространяющегося в двухфазной среде с учетом лучистого переноса в зависимости от размеров расчетной ячейки воспроизводящих ширину фронта, $L_f / \Delta$. Как видно из Рис. П1 сходимость в случае двухфазной среды с переносом теплового излучения микрочастицами достигается при том же разрешении, что и для решения для обычного пламени без частиц, при разрешении фронта пламени 24 ячейками с размером 0.01мм. Полученный результат показывает, что численный алгоритм, используемый для решения задач горения в классической постановке, с той же степенью точности применим и для решения задач горения дисперсных сред с учётом лучистого переноса.

Рисунок П2 показывает результаты тестов на сходимость по скорости горения для трех различных начальных температур свежей смеси (293К, 600К и 1000К). Размер расчетных ячеек дан в размерных единицах. Закрашенные символы соответствуют разрешению, обеспечивающему сходимость решения. Из Рис. П2 видно, что при увеличении начальной температуры смеси и соответствующем увеличении скорости горения, а так же при уменьшении ширины фронта реакции, требуется существенное увеличение пространственного разрешения до $10^{-3}$ мм для сходимости решения задачи о воспламенении за счет поглощения энергии на неоднородностях распределения частиц. Полученная граница сходимости решений, соответствующих различным фоновым температурам, определяет выбор расчётных параметров при численном воспроизведении исследуемых в работе режимов горения, в частности, распространение спонтанной волны горения вдоль градиента температуры, развитие нестационарного горения при повышенных температурах и давлениях за отходящей ударной волной, взаимодействие фронтов горения и ударных волн, и формирование детонации [24-28].

Рис. П3 показывает значения скоростей горения, полученные в экспериментах [48] и рассчитанные при разных значениях начальной температуры смеси. Так как информация об особенностях развития процесса при проведении экспериментов ограничена, в частности нет количественного описания процесса инициирования волны горения, а следовательно нет информации о волне сжатия, генерируемой в области подвода энергии, то вычисленные значения скорости горения на Рис. П3 даны с погрешностью, учитывающей неопределенность температуры за отходящей из области воспламенения волной сжатия. Сравнения экспериментальных и расчётных результатов, представленных на Рис. П3, показывает хорошее воспроизведение в расчётах скорости волны горения в широком интервале изменения фоновой температуры, что позволяет ожидать корректное воспроизведение на вычислительном эксперименте смены режимов горения при развитии нестационарных переходных процессов, включая переход горения в детонацию.

Процессы, исследуемые в настоящей работе, включают также генерацию ударных и детонационных волн. Корректность описания ударно-волновых процессов при использовании принятого вычислительного алгоритма исследовалось как в работе [21], так и в более поздних публикациях [49]. Успешное применение метода к решению задач горения и детонации продемонстрировано в работах авторов [24-28]. Сходимость метода при решении задач горения представлена в работе [15], где вычислительный алгоритм проверялся при решении задачи о спонтанной волне реакции на начальном градиенте температуры.

В рассматриваемых задачах, принципиальным является корректное описание зажигания газовой горючей смеси, то есть правильное определение условий перехода от эндотермической к экзотермической стадии реакции. Относительно простая редуцированная модель химической кинетики [16] состоит из 19 реакций для 8 компонент, однако, как и другие современные редуцированные схемы эта схема достаточно хорошо воспроизводит процесс зажигания. Поскольку целью настоящей работы была качественная демонстрация принципиально новых физических процессов, связанных с влиянием излучения на режимы горения, то небольшие различия, присущие используемой в работе и всем другим детальным схемам химических реакций в описании горения в широком диапазоне параметров смеси, не представляются принципиальными. В то же время следует отметить, что схема с одноступенчатой химической реакцией, в которой нет порога зажигания, и реакция начинается при любой температуре, даже качественно не отражают реальный процесс воспламенения горючей смеси.

**Подписи к рисункам**

Рисунок 1. Эволюция температуры газа перед фронтом водород-кислородного пламени (на расстоянии 2мм от мгновенного положения фронта) при различных длинах свободного пробега излучения. Время на оси абсцисс в единицах $L/U_{f0}$.

Рисунок 2. Профили температуры установившиеся в стационарном режиме перед фронтом водород-кислородного пламени при различных длинах свободного пробега излучения. Тонкой линией показан профиль температуры, соответствующий ламинарному водород-кислородного пламени в отсутствии микрочастиц.

Рисунок 3. Эволюция скоростей горения водород-кислородной смеси при различных длинах свободного пробега излучения, приведенных к нормальной ламинарной скорости горения в газе без частиц. Время на оси абсцисс в единицах $L/U_{f0}$.

Рисунок 4. Изменение ширины фронта водород-кислородного пламени в зависимости от длины пробега излучения.

Рисунок 5. Схематическая картина распространения пламени в газовой смеси с неоднородным пространственным распределением микрочастиц. 1 – область горячих продуктов горения за фронтом пламени, 2 – область свободного распространения излучения в среде с малой концентрацией микрочастиц, 3 – область поглощения излучения локально сконцентрированными микрочастицами.

Рисунок 6. Профили температуры газа (а) и плотности частиц (б) вблизи границы более плотного облака частиц перед фронтом пламени в ходе его нагрева излучением. Профили приведены на различные моменты времени с шагом 50мкс. $N_p=2.5 \cdot 10^7$, $r_p=1$мкм.

Рисунок 7. Эволюция профилей температуры газа (средняя часть рисунка) и давления (нижняя часть) в процессе формирования волны медленного горения вблизи границы облака частиц перед фронтом пламени. $t_0=900$мкс, $\Delta t=50$мкс. Верхняя часть рисунка показывает начальное распределение частиц (штриховая линия) и распределение частиц на момент времени $t_0$ (сплошная).

Рисунок 8. Эволюция профилей температуры газа (средняя часть рисунка) и давления (нижняя часть) в процессе формирования волны медленного горения и ударной волны вблизи границы облака частиц. $t_0=1650$мкс, $\Delta t=50$мкс. Верхняя часть рисунка показывает начальное распределение частиц (штриховая линия) и распределение частиц на момент времени $t_0$ (сплошная).

Рисунок 9. Эволюция профилей температуры газа (средняя часть рисунка) и давления (нижняя часть) в процессе образования волны детонации вблизи границы облака частиц. $t_0$=4980мкс, Δt=4мкс. Верхняя часть рисунка показывает начальное распределение частиц (штриховая линия) и распределение частиц на момент времени $t_0$ (сплошная).

Рисунок П1. Нормальная скорость горения $U_f$, адиабатическая температура продуктов горения $T_b$ и коэффициент расширения $\theta = \rho_u / \rho_b$ для ламинарного пламени в стехиометрической водород-кислородной смеси при нормальных условиях $T_0 = 300K$, $p_0 = 1atm$ для различных размеров расчетных ячеек. $L_f$ - ширина фронта, $\Delta$ - размер расчетной ячейки, индекс 'c' соответствует сходимости решения. Кружки - значения скорости в газе без частиц, заполненные кружки - значения скорости пламени в газе с частицами.

Рисунок П2. Тесты на сходимость для различных начальных температур перед фронтом пламени. Затененные точки показывают область сходимости.

Рисунок П3. Скорость горения в стехиометрической водород-кислородной смеси в зависимости от начальной температуры смеси. Сплошная линия (1) показывает экстраполяцию экспериментальных данных [48] (показаны квадратами); ромбы - вычисленные значения. Также показаны отклонения, связанные с ростом температуры за волной сжатия.

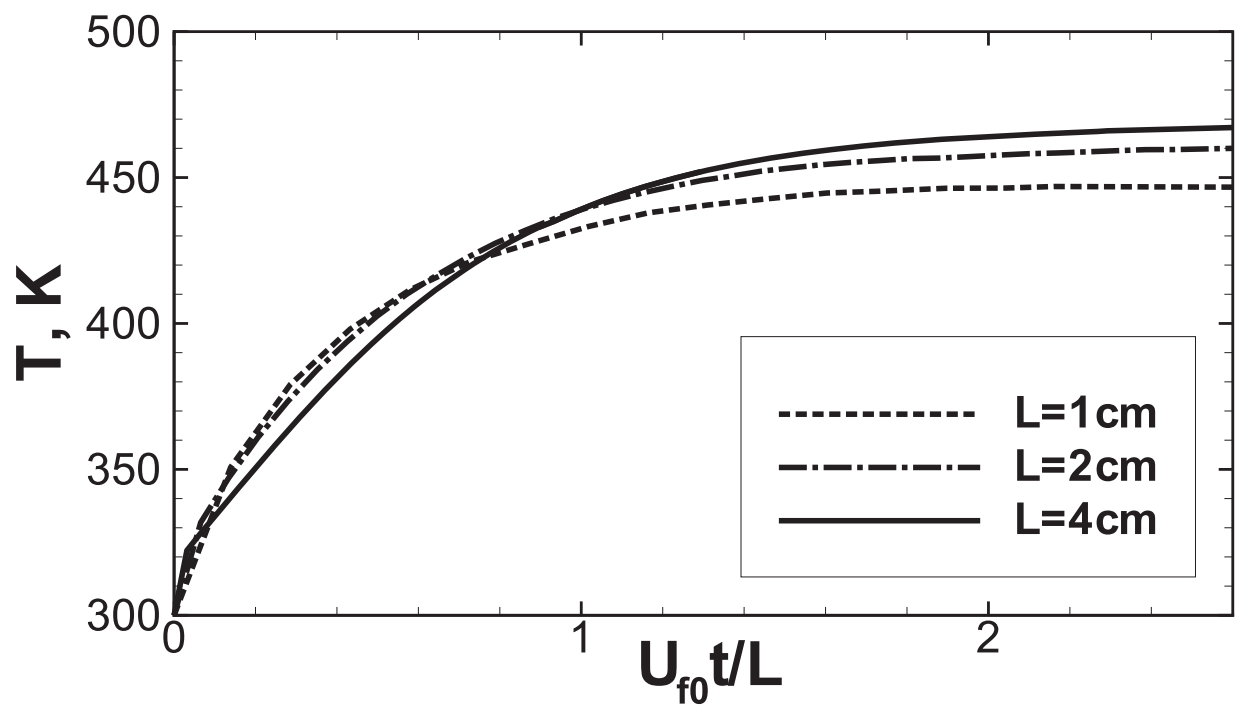

Рисунок 1. Эволюция температуры газа перед фронтом водород-кислородного пламени (на расстоянии 2мм от мгновенного положения фронта) при различных длинах свободного пробега излучения. Время на оси абсцисс в единицах $L/U_{f0}$.

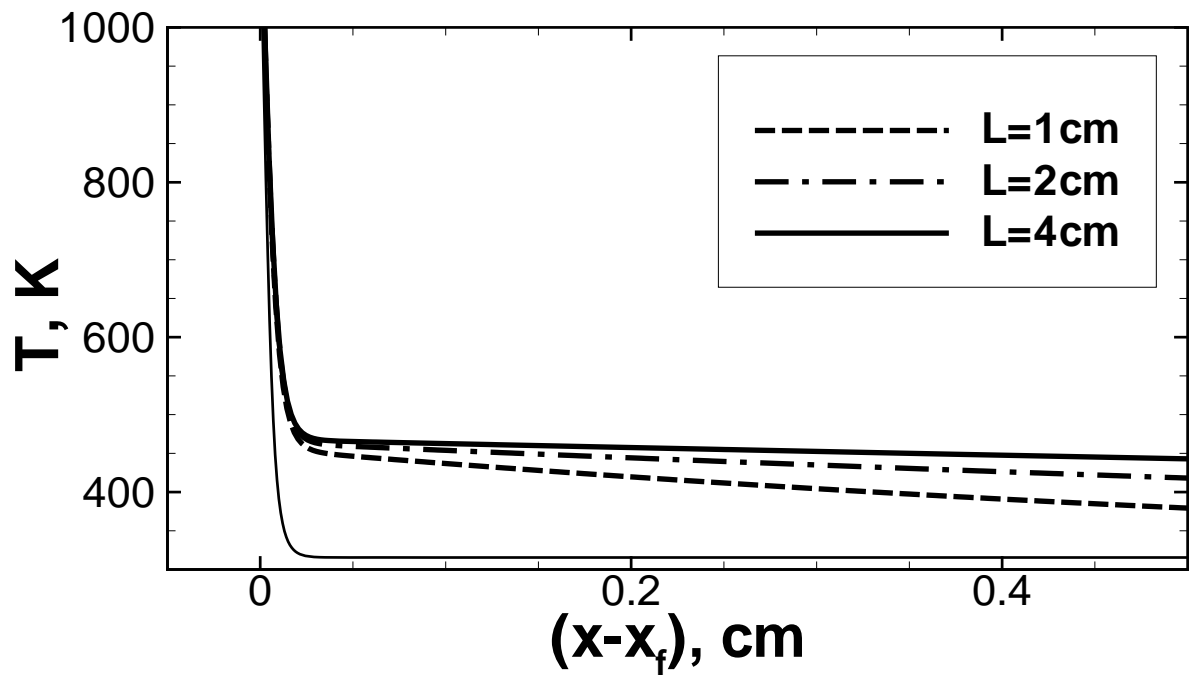

Рисунок 2. Профили температуры установившиеся в стационарном режиме перед фронтом водород-кислородного пламени при различных длинах свободного пробега излучения. Тонкой линией показан профиль температуры, соответствующий ламинарному водород-кислородного пламени в отсутствии микрочастиц.

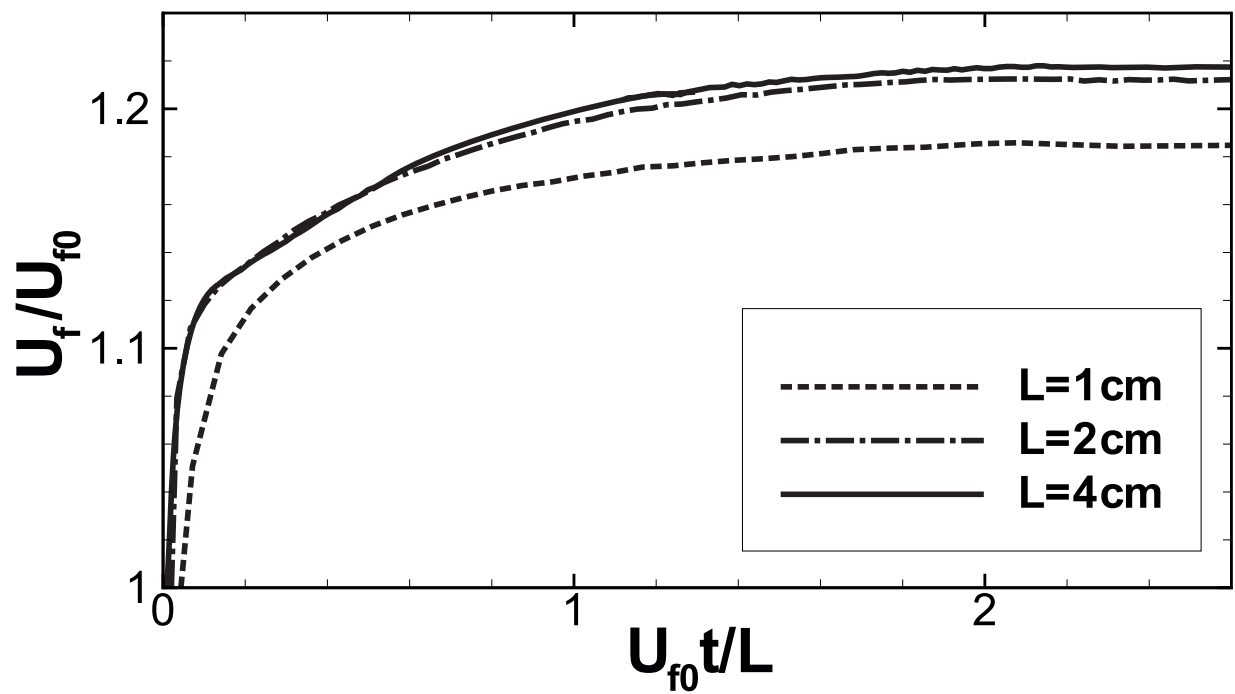

Рисунок 3. Эволюция скоростей горения водород-кислородной смеси при различных длинах свободного пробега излучения, приведенных к нормальной ламинарной скорости горения в газе без частиц. Время на оси абсцисс в единицах $L/U_{f0}$.

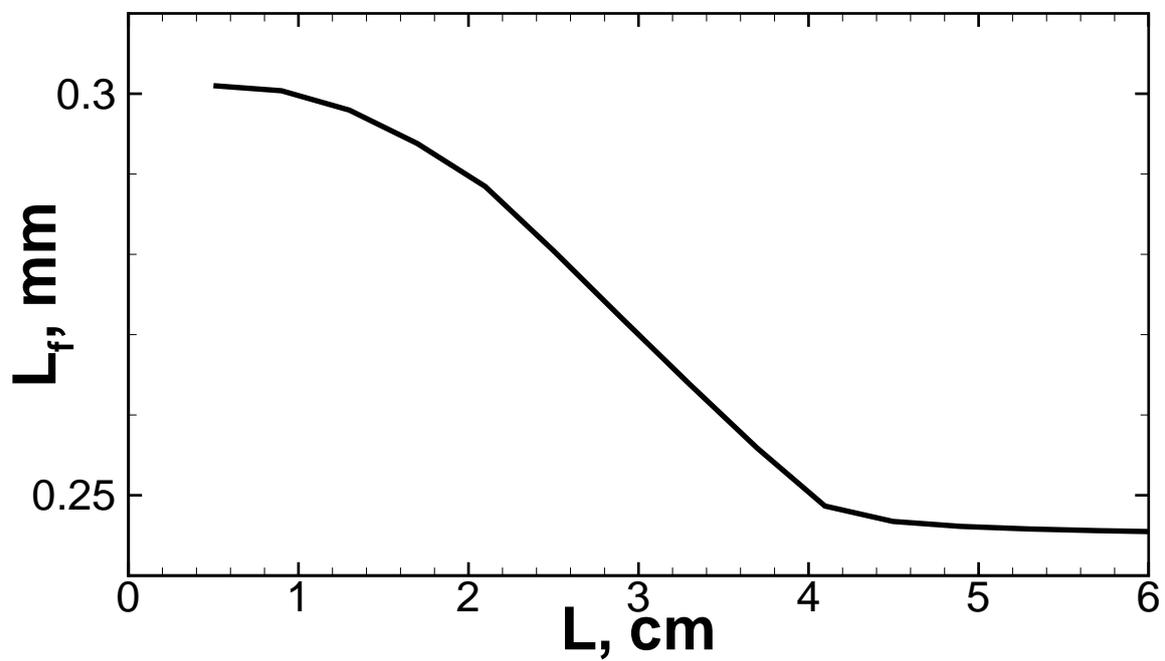

Рисунок 4. Изменение ширины фронта водород-кислородного пламени в зависимости от длины пробега излучения.

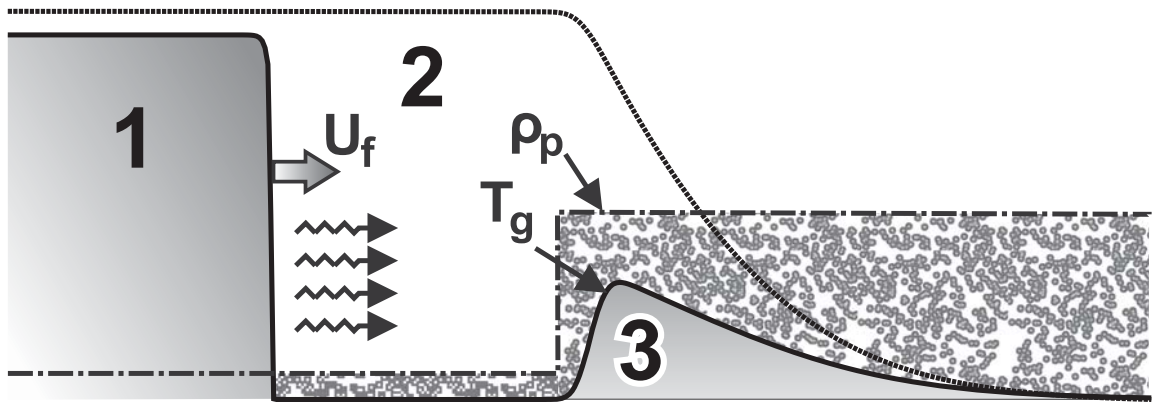

Рисунок 5. Схематическая картина распространения пламени в газовой смеси с неоднородным пространственным распределением микрочастиц. 1 – область горячих продуктов горения за фронтом пламени, 2 – область свободного распространения излучения в среде с малой концентрацией микрочастиц, 3 – область поглощения излучения локально сконцентрированными микрочастицами.

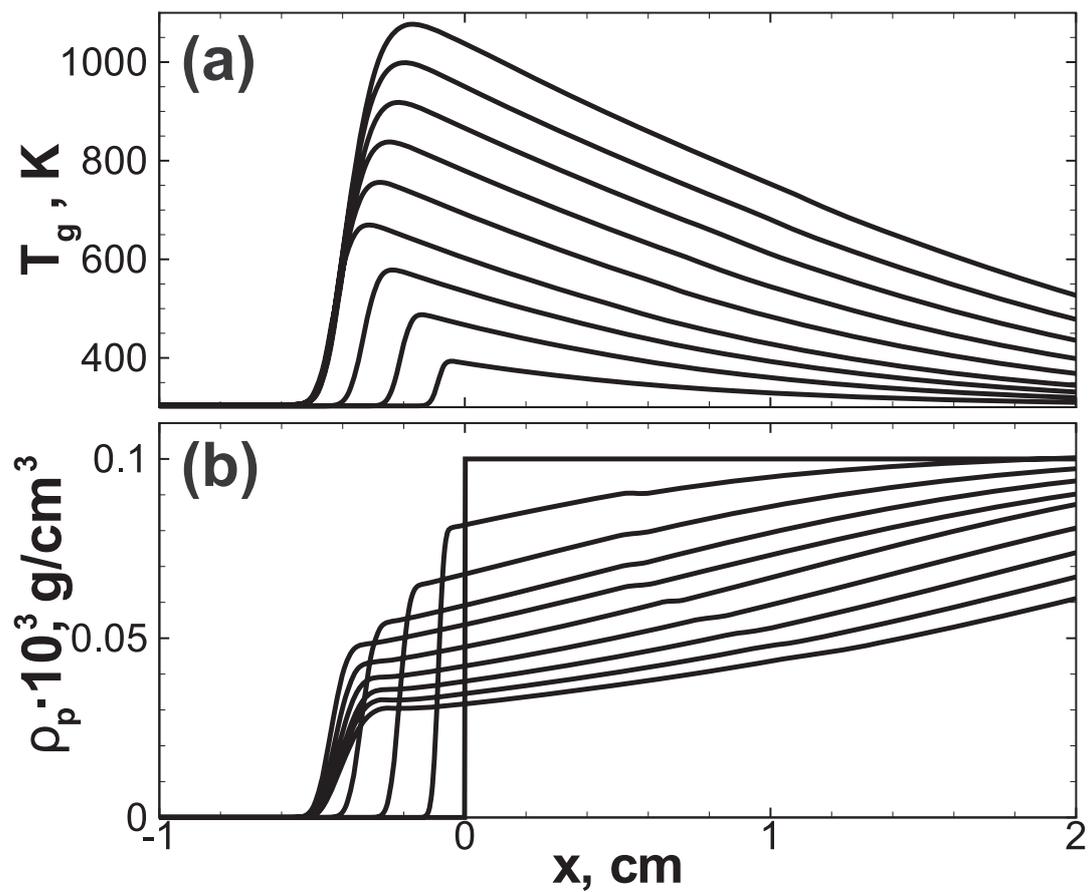

Рисунок 6. Профили температуры газа (а) и плотности частиц (б) вблизи границы более плотного облака частиц перед фронтом пламени в ходе его нагрева излучением. Профили приведены на различные моменты времени с шагом 50мкс. $N_p=2.5 \cdot 10^7$, $r_p=1$мкм.

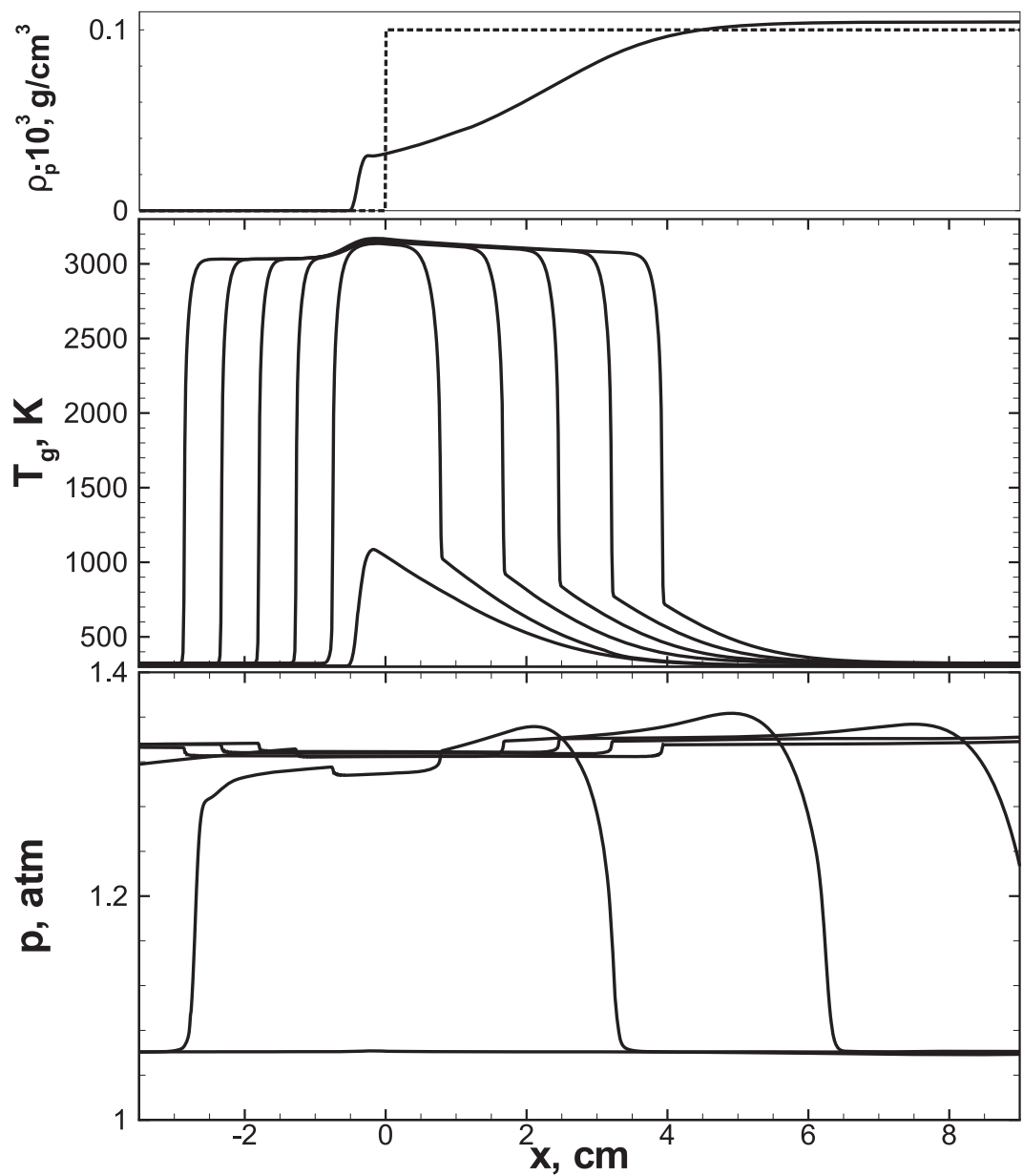

Рисунок 7. Эволюция профилей температуры газа (средняя часть рисунка) и давления (нижняя часть) в процессе формирования волны медленного горения вблизи границы облака частиц перед фронтом пламени. $t_0$=900мкс, $\Delta t$=50мкс. Верхняя часть рисунка показывает начальное распределение частиц (штриховая линия) и распределение частиц на момент времени $t_0$ (сплошная).

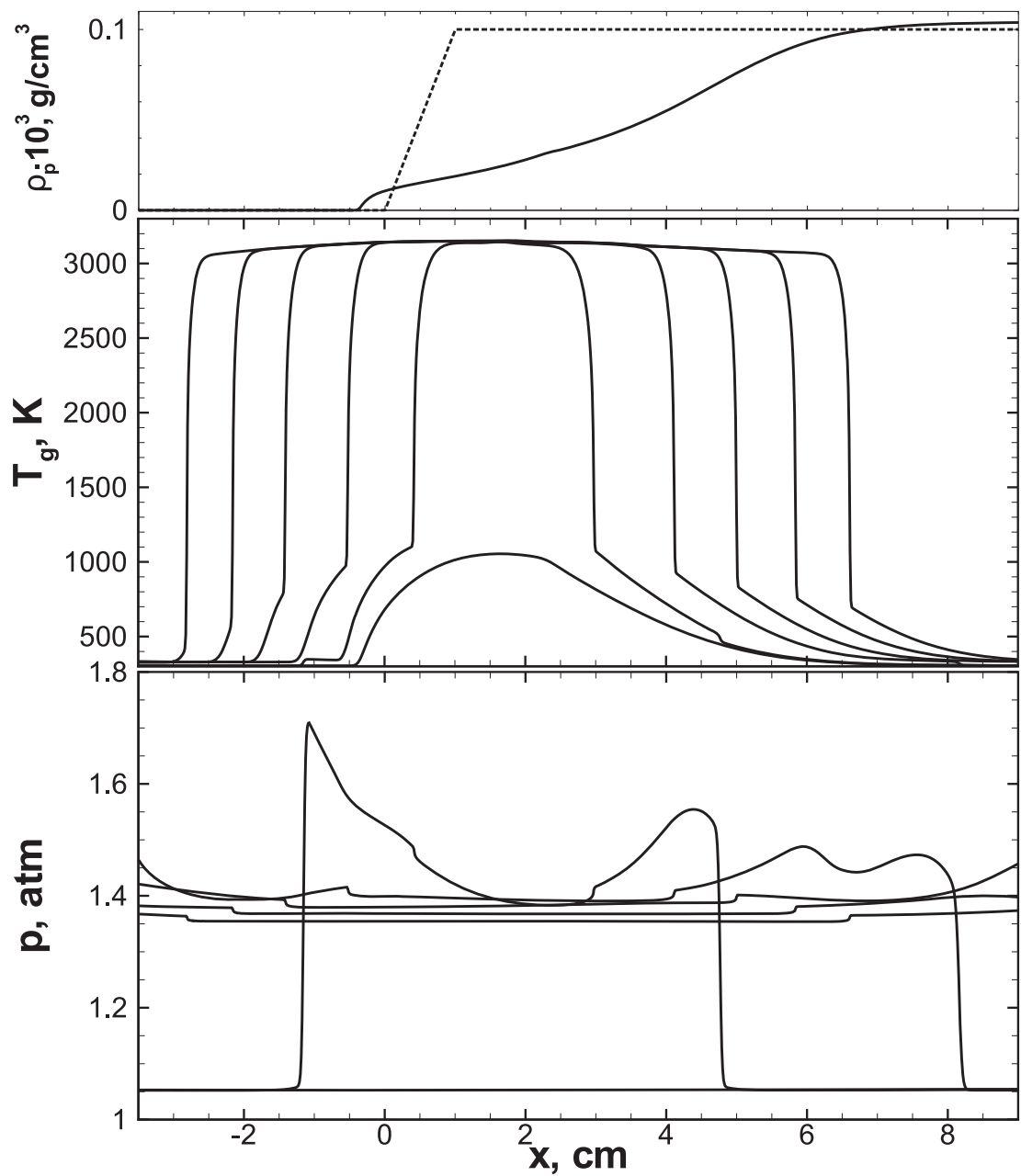

Рисунок 8. Эволюция профилей температуры газа (средняя часть рисунка) и давления (нижняя часть) в процессе формирования волны медленного горения и ударной волны вблизи границы облака частиц. $t_0=1650$мкс, $\Delta t=50$мкс. Верхняя часть рисунка показывает начальное распределение частиц (штриховая линия) и распределение частиц на момент времени $t_0$ (сплошная).

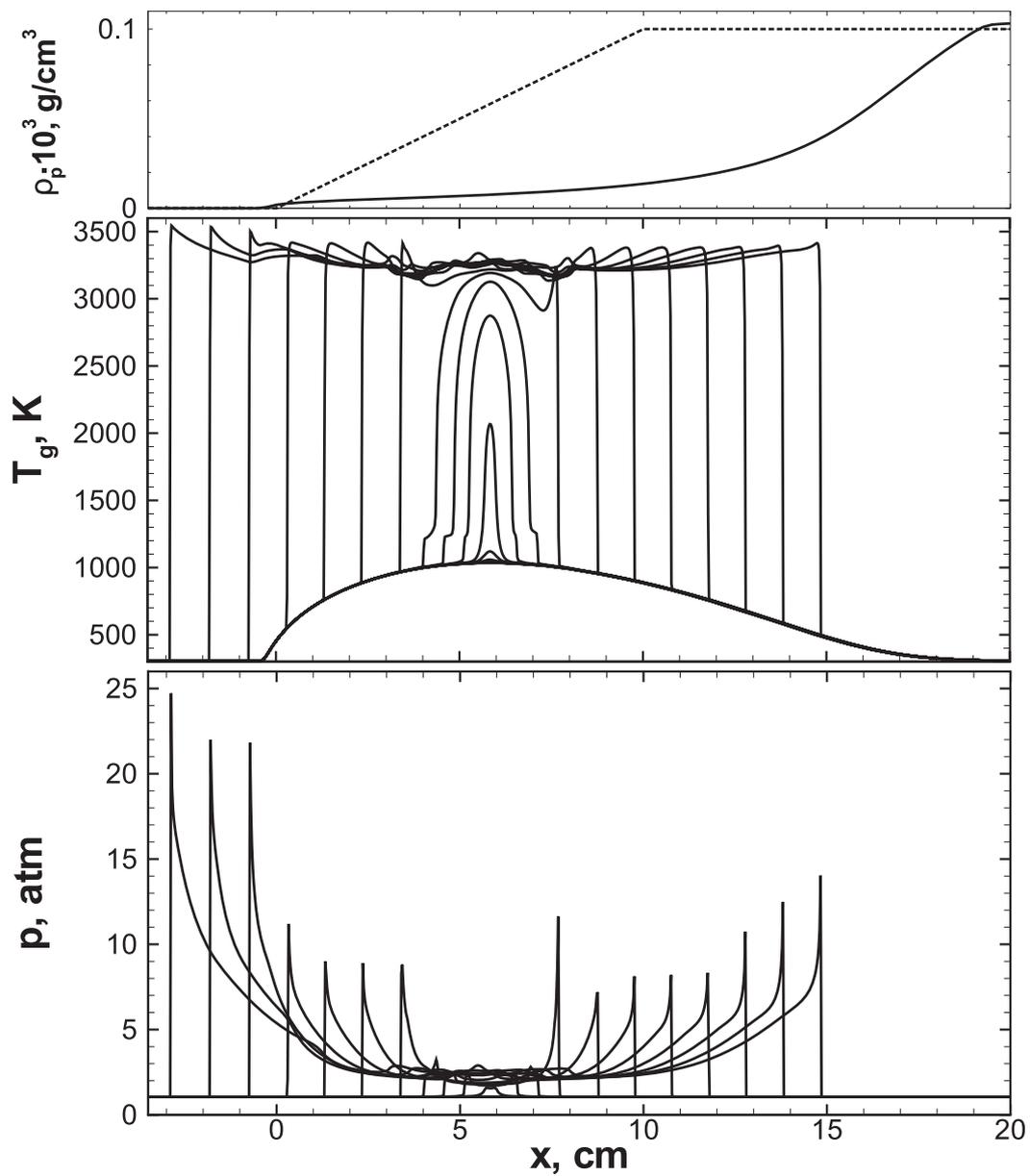

Рисунок 9. Эволюция профилей температуры газа (средняя часть рисунка) и давления (нижняя часть) в процессе образования волны детонации вблизи границы облака частиц. $t_0$=4980мкс, $\Delta t$=4мкс. Верхняя часть рисунка показывает начальное распределение частиц (штриховая линия) и распределение частиц на момент времени $t_0$ (сплошная).

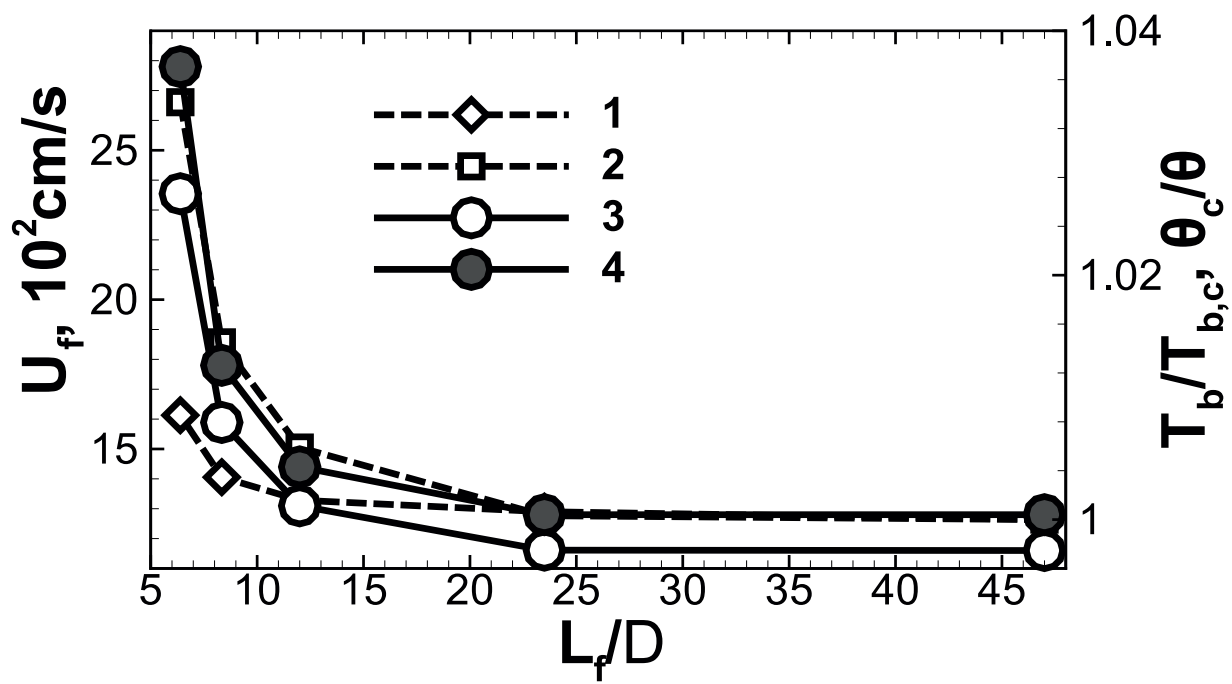

Рисунок П1. Нормальная скорость горения $U_f$, адиабатическая температура продуктов горения $T_b$ и коэффициент расширения $\theta = \rho_u/\rho_b$ для ламинарного пламени в стехиометрической водород-кислородной смеси при нормальных условиях $T_0 = 300K$, $p_0 = 1atm$ для различных размеров расчетных ячеек. $L_f$ - ширина фронта, $\Delta$ - размер расчетной ячейки, индекс 'c' соответствует сходимости решения. Кружки - значения скорости в газе без частиц, заполненные кружки - значения скорости пламени в газе с частицами.

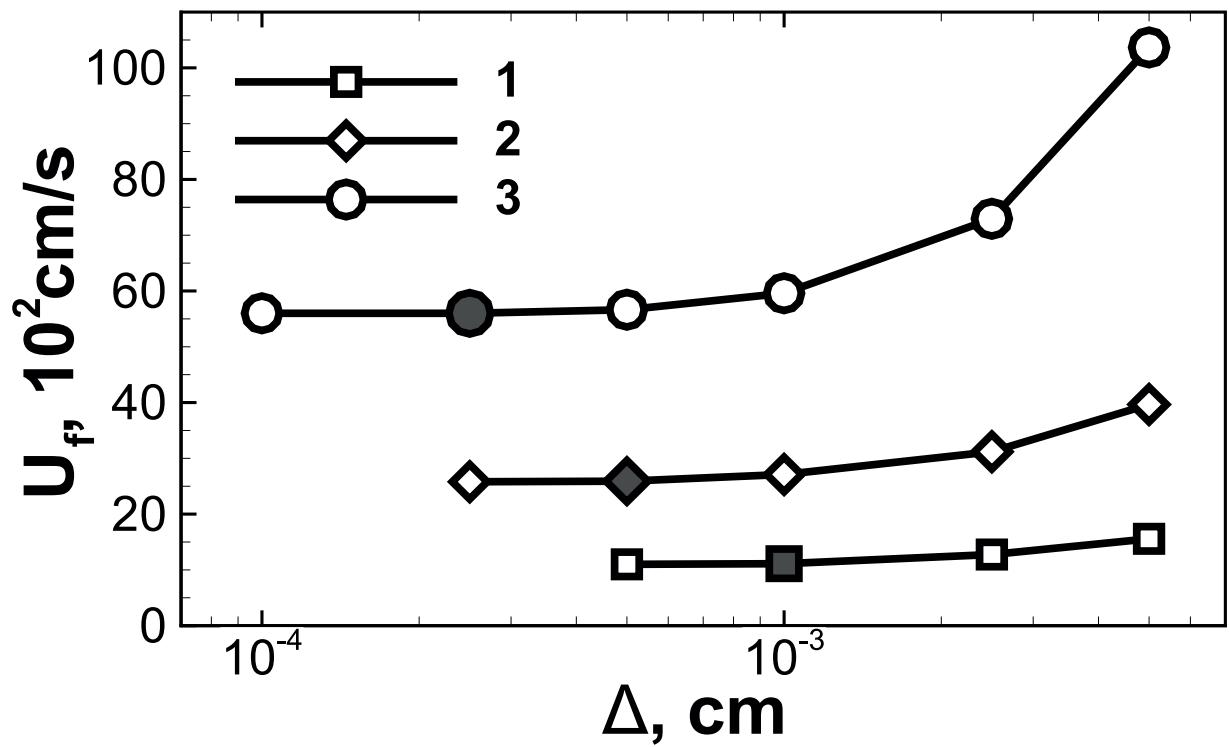

Рисунок П2. Тесты на сходимость для различных начальных температур перед фронтом пламени. Затененные точки показывают область сходимости.

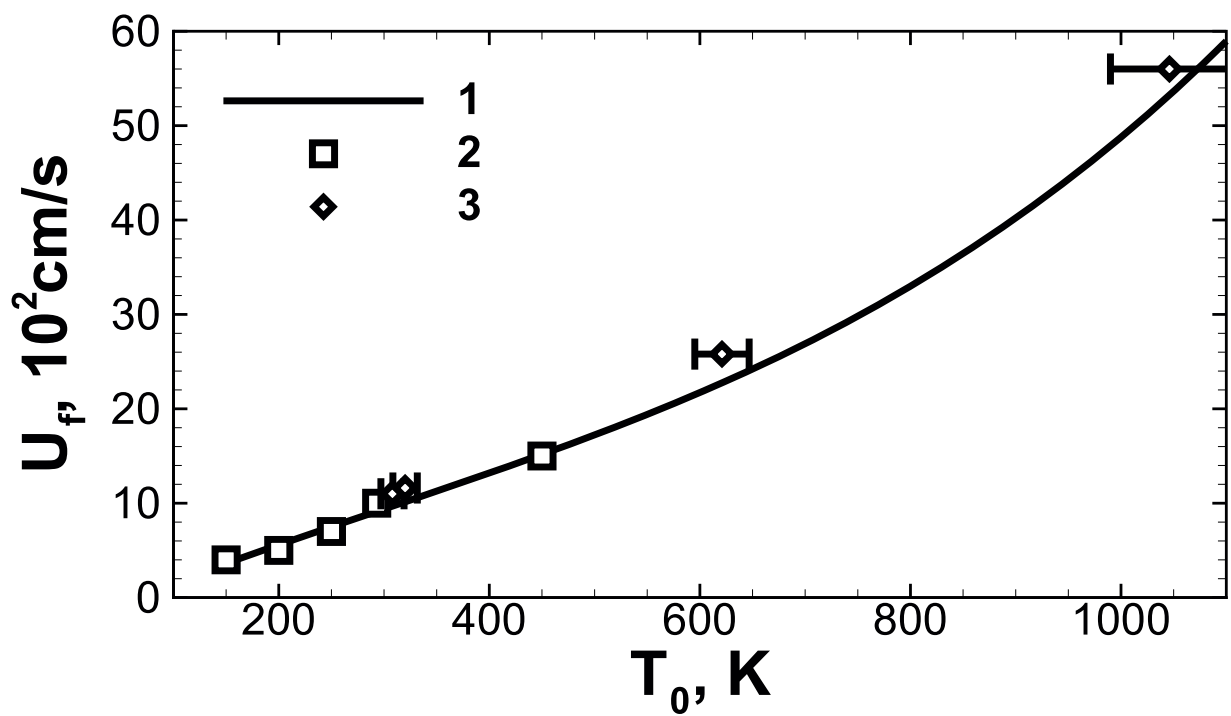

Рисунок П3. Скорость горения в стехиометрической водород-кислородной смеси в зависимости от начальной температуры смеси. Сплошная линия (1) показывает экстраполяцию экспериментальных данных [48] (показаны квадратами); ромбы - вычисленные значения. Также показаны отклонения, связанные с ростом температуры за волной сжатия.